\documentclass[12pt,preprint]{aastex}
\begin{document}

\title{\vspace*{-1.0in}The Secular Evolution of a 
  Close Ring--Satellite System:\\
  The Excitation of Spiral Bending Waves at a 
  Nearby Gap Edge\vspace*{0.25in}
}

\author{
  Joseph M. Hahn
}
\affil{
  Space Science Institute\\
  10500 Loring Drive\\
  Austin, TX, 78750\\
  email: jhahn@spacescience.org\\
  phone: 512-291-2255\vspace*{0.25in}
}

\author{Accepted for publication in the {\it Astrophysical Journal}\\
  April 30, 2007
  \vspace{0.25in}
}

\begin{abstract}

The secular perturbations exerted by an inclined satellite orbiting in a gap in a 
broad planetary ring tends to excite the inclinations of the nearby ring particles,
and the ring's self--gravity can allow that disturbance
to propagate away in the form of a spiral bending wave.
The amplitude of this spiral bending wave is determined, as well
as the wavelength, which shrinks as the waves propagate outwards
due to the effects of the central planet's oblateness.
The excitation of these bending waves also damps the satellite's 
inclination $I$. This secular $I$ damping 
is also compared to the inclination excitation that is
due to the satellite's many other vertical resonances in the ring, and the condition
for inclination damping is determined. The secular $I$ damping is likely
responsible for confining the orbits of Saturn's two known
gap--embedded moons, Pan and Daphnis, to the ring plane.

\end{abstract}


\section{Introduction}
\label{intro}

Secular gravitational
perturbations can play a significant
role in determining the global structure and the
long--term evolution of a
disk--companion system.  A well--known example is the circumstellar dust disk at 
$\beta$ Pictoris, whose broad but gentle
warp is thought to be due to the secular gravitational
perturbations exerted by an unseen
planetary system \citep{Metal97}. Secular perturbations from an eccentric planet
can also make a dust--disk appear lopsided, too \citep{Wetal99}.
Secular perturbations are those forces
that are due to the time--independent part of a companion's  gravitational potential,
and such perturbations are equivalent to the gravitational forces that arise when a
perturber's mass is spread about its orbital ellipse \citep{MD99}.
Consequently, the long--term secular evolution of a disk--companion system is
conveniently modeled by treating it as a system of gravitating
rings \citep{H03}, an approach that will also be employed here.

When the perturber's orbit is eccentric or inclined, 
its secular perturbations can excite
the orbital eccentricities or inclinations of the disk particles,
as well as cause the orbits of the disk particles to precess over time.
Large eccentricities $e$ or inclinations $I$ can also be excited at
a secular resonance, which is a site where
the disk matter precesses in sync with one of the eigenfrequencies
that describe the perturber's precession. However, substantial 
$e$'s and $I$'s can also be excited elsewhere at
non--resonant sites in the disk, with greater excitation occurring nearer
the perturber. In fact, it is these
non--resonant secular perturbations of the disk that are the focus of this study.

If the disk has internal forces, such as pressure or self--gravity, then
those internal forces can also transmit the companion's 
disturbances across the disk.
For instance, an inclined planet orbiting in a circumstellar
gas disk can excite a global warp that is facilitated by the disk's
internal pressure \citep{LO01}.
And if an eccentric companion inhabits a gap in the gas disk,
its secular perturbations can launch a density wave at the gap edge
having such a long wavelength
that a global standing wave emerges  \citep{GS03}.
But if the disk is instead gravity--dominated, then the companion can launch
spiral density or spiral bending waves
at its secular resonances in the disk \citep{WH98, WH03}.
These phenomena are also relevant to studies of extra--solar planets,
since any dissipation in the disk facilitates a transfer of angular momentum
between the disk and the companion in a manner that tends to drive its eccentricity or
inclination to zero.

The following will examine the secular evolution of a related system:
of a small satellite that inhabits a narrow gap in a broad planetary ring, both of which
are orbiting an oblate central planet. It will be shown
below that an inclined satellite can launch a spiral bending wave that propagates
outwards and away from the gap's outer edge. The amplitude
and wavelength of this spiral bending wave is assessed below, as well as the
rate at which this wave--action damps the satellite's inclination. 
This secular inclination--damping mechanism is then compared
to the inclination excitation
that is due to the satellite's many other vertical resonances in the ring
(e.g., \citealt{BGT84}). We then quantify when this secular damping dominates over
the resonant excitation, and show that this secular interaction is likely
responsible for confining Saturn's two known
gap--embedded moons, Pan and Daphnis, to the ring plane.

\section{Equations of motion}
\label{EOM}

Begin by considering
a planetary ring that is perturbed by a single satellite, 
with both orbiting an oblate planet. 
To assess the disturbance that the satellite might launch in this ring,
the Lagrange planetary equations will be used; they
give the rates at which a ring particle's orbital inclination $I$
and longitude of ascending node $\Omega$ varies with time $t$:
\begin{equation}
  \label{Idot}
  \dot{I}\simeq-\frac{1}{na^2I}\frac{\partial R}{\partial\Omega}
    \quad\mbox{and}\quad
  \dot{\Omega}\simeq\frac{1}{na^2I}\frac{\partial R}{\partial I},
\end{equation}
where $R$ is the disturbing function for a ring particle having a semimajor axis
$a$ and mean motion $n\simeq\sqrt{GM/a^3}$, 
where $G$ is the gravitation constant
and $M$ is the mass of the central planet \citep{MD99},
and all inclinations are small, $I\ll1$. The total disturbing
function for a ring particle is $R=R_{disk}+R_{sat}+R_{obl}$,
where the three terms account for the gravitational perturbations
that are due to ring's gravity (which we
treat here as a broad disk), the satellite's perturbations, and that due to the
planet's oblate figure. The particle's
equations of motion is thus the sum of three parts:
\begin{equation}
  \dot{I}=\left.\dot{I}\right|_{disk}+\left.\dot{I}\right|_{sat}
    \quad\mbox{and}\quad
    \dot{\Omega}=\left.\dot{\Omega}\right|_{disk}+
    \left.\dot{\Omega}\right|_{sat}+\left.\dot{\Omega}\right|_{obl},
\end{equation}
noting that oblateness does not alter inclinations. And because we are only dealing
with the system's secular perturbations, the semimajor axes of all bodies are constant
\citep{BC61}.

The amplitude of a spiral bending wave that is in a 
steady--state does not vary with time,
so the disk inclinations obey
$\dot{I}(a)=0$ throughout the disk. A persistent spiral pattern must also
rotate with a constant angular velocity $\omega$, so
\begin{eqnarray}
  \label{ss}
  \left.\dot{I}\right|_{disk}&=&-\left.\dot{I}\right|_{sat}\\
  \label{omega}
  \mbox{and}\quad\omega&=&\left.\dot{\Omega}\right|_{disk}+
    \left.\dot{\Omega}\right|_{sat}+\left.\dot{\Omega}\right|_{obl}
    =\mbox{ constant}.
\end{eqnarray}
The following subsection will use the first equation to solve for the wave
amplitude $I(a)$ throughout the disk. The next subsection
will then use the other equation to solve
for the bending waves' dispersion relation $\omega(k)$, which in turn
provides the wavenumber $k$ of the spiral bending wave, and the wave's 
radial velocity.

\subsection{wave amplitude}
\label{amplitude}

Begin by examining how the planetary ring perturbs itself.
The ring is to be regarded as a broad disk that is composed 
of many narrow, concentric annuli.
Each annulus has mass $\delta m(a)$, 
inclination $I(a)$, and longitude
of ascending node $\Omega(a)$, all of which are to be regarded as functions
of the rings' semimajor axes $a$.
For the moment we will assume that all rings
are circular, noting that we will deal with the system's eccentricity evolution
in a followup study \citep{H07}.
Suppose that the annulus at $a$ is perturbed by another annulus of mass $\delta m'$
and radius $a'$; the disturbing function for the perturbed annulus is
\begin{equation}
  \label{deltaR0}
  \delta R=-\frac{G\delta m'}{4a}\alpha 
    \tilde{b}^{(1)}_{3/2}(\alpha)\left[\frac{1}{2}I^2-II'\cos(\Omega-\Omega')  \right],
\end{equation}
where $a, I, \Omega$ are the orbit elements of the perturbed annulus,
and the primed quantities refer to the perturbing annulus  \citep{H03}. 
The softened Laplace coefficient appearing in the above is
\begin{equation}
  \label{lc}
  \tilde{b}^{(1)}_{3/2}(\alpha)=\frac{2}{\pi}\int_0^\pi
    \frac{\cos(\varphi)d\varphi}
    {[(1+\alpha^2)(1+\mathfrak{h}^2)-2\alpha\cos\varphi]^{3/2}},
\end{equation}
and it is a function of the semimajor axis ratio $\alpha=a'/a$, 
where $\mathfrak{h}=h/a\ll1$
is the disk's vertical scale height $h$ in units of semimajor axis $a$.
Note that when the disk is infinitesimally thin, $\mathfrak{h}=0$,
and the disturbing function $\delta R$ is equivalent to that due to a point--mass
$\delta m'$ (e.g., \citealt{BC61}).

\begin{figure}
\epsscale{1.0}
\plotone{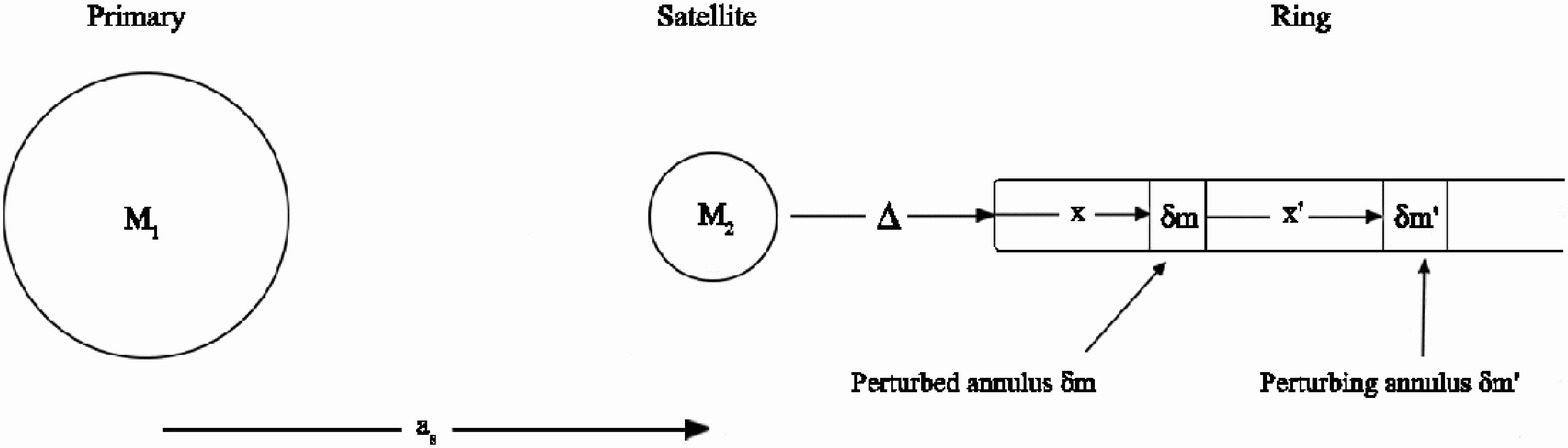}
\figcaption{
  \label{geometry}
  A schematic showing the geometry of the ring--satellite system, seen edge-on. A
  satellite of mass $m_s$ and semimajor axis $a_s$
  orbits interior to a broad planetary ring that extends to infinity.
  The satellite's distance from the ring's inner edge is $\Delta$ in units of the
  satellite's semimajor axis $a_s$. A perturbed annulus in the ring has mass $\delta m$,
  and it lies a fractional distance $x$ away from the ring's inner edge,
  while the perturbing ring has mass $\delta m'$ and lies a fractional distance $x'$
  from the perturbed ring.
}
\end{figure}

It will be convenient to replace the ring mass
$\delta m'$ with $2\pi\sigma'a'da'$, where $\sigma'=\sigma(a')$ 
is the mass surface density of the annulus of radius $a'$ and radial width $da'$. 
We will also write its semimajor axis as $a'=a(1+x')$,
where $x'=(a'-a)/a=\alpha-1$ is the fractional distance 
between the perturbing ring $a'$ and
the perturbed ring $a$. 
For the moment we will consider a one-sided disk---one that orbits wholly
exterior to the satellite, where
$\Delta$ is the fractional distance between the satellite's orbit
and the disk's inner edge; the geometry is sketched in Fig.\ \ref{geometry}. 
The disturbing function for ring $a$
due to perturbations from ring $a'$ can now be written as
\begin{equation}
  \label{deltaR}
  \delta R=-\frac{1}{2}\mu_d'(na)^2 \tilde{b}^{(1)}_{3/2}(x')
    \left[\frac{1}{2}I^2-II'\cos(\Omega-\Omega')\right]dx'
\end{equation}
where $\mu_d'\equiv\pi\sigma'a'^2/M$ is the ring's so--called normalized disk mass,
$dx'=da'/a$ is the perturbing ring's fractional width, and $\tilde{b}^{(1)}_{3/2}(x')$
is shorthand for Eqn.\ (\ref{lc}) evaluated at $\alpha=1+x'$.  Then
according to Eqn.\ (\ref{Idot}), ring $a'$ will alter the inclination of ring $a$ at the 
rate
\begin{equation}
  \label{delta-dot-I}
  \delta\dot{I}=-\frac{1}{na^2I}\frac{\partial(\delta R)}{\partial\Omega}
    =\frac{1}{2}\mu_d'n \tilde{b}^{(1)}_{3/2}(x')I'\sin(\Omega-\Omega')dx'.
\end{equation}

\subsubsection{ring--disk evolution}

The total rate at which the
entire disk alters the inclination of ring $a$ is the above with $x'$ integrated across the
disk, from $-x$ to $+\infty$ (see Fig.\ \ref{geometry}), so
\begin{equation}
  \label{Idot_disk0}
  \left.\dot{I}\right|_{disk}=\int_{disk}\delta\dot{I}=
    \frac{1}{2}n\int_{-x}^\infty
    \mu_d'(x')\tilde{b}^{(1)}_{3/2}(x')I'(x')\sin(\Omega-\Omega')dx'.
\end{equation}

As one might expect, this integral is dominated by the contributions from
nearby annuli that lie a small distance $x'$ away. 
In the $|x'|\ll1$ limit, the softened Laplace coefficient is
\begin{equation}
  \label{b_approx}
  \tilde{b}^{(1)}_{3/2}(x')\simeq\frac{2}{\pi(x'^2+2\mathfrak{h}^2)}
\end{equation}
\citep{H03}. Because of the steep dependence of $\tilde{b}^{(1)}_{3/2}$ on $x'$,
we can replace the inclination $I'(x')$ and disk mass $\mu_d'(x')$ with their values
evaluated at the perturbed ring at $x'=0$, so
$I'\simeq I$ and $\mu_d'\simeq\mu_d=\pi\sigma a^2/M$,
and also pull them out of the integral so that
\begin{equation}
  \label{Idisk0}
  \left.\dot{I}\right|_{disk}\simeq
    \frac{\mu_dIn}{\pi}\int^\infty_{-x}
    \frac{\sin[\Omega-\Omega'(x')]}{x'^2+2\mathfrak{h}^2}dx'.
\end{equation}
Most of the contributions to this integral will be due to nearby annuli that lie
a wavelength $\lambda\simeq2\pi/|k|$ away, where $k$ is the wavenumber of the
spiral bending wave.

A spiral wave has a wavenumber $k(a)=-\partial\Omega/\partial a$
(Eqn.\ \ref{k_appendix}), so the $\Omega-\Omega'$ in the above is
\begin{equation}
  \label{wavenumber_exact}
  \Omega(a)-\Omega'(a')=-\int_{a'}^a k(r)dr.
\end{equation}
In general, the wavenumber $k(a)$ will vary with semimajor axis $a$. 
However, considerable progress can be made if we assume that
$k$ is constant over a wavelength,
so $\Omega-\Omega'\simeq-k(a-a')=kax'$. Then Eqn.\ (\ref{Idisk0}) becomes
\begin{equation}
  \label{Idot_disk}
  \left.\dot{I}\right|_{disk}\simeq\frac{1}{\pi}A_H(|k|ax)\mu_dIkan
\end{equation}
after replacing the $x'$ integration variable with $y=|k|ax'$, and noting that
Eqn.\ (\ref{Idisk0}) is odd in $y$. In the above, the function $A(z)$ 
is a dimensionless measure of the warped disk's perturbation of itself:
\begin{equation}
  \label{A}
  A_H(z)\equiv\int_z^\infty\frac{\sin y}{y^2+H^2}dy,
\end{equation}
where $z=|k|ax$ is the distance from the ring--edge in units of
$2\pi$ wavelengths, and the dimensionless wavenumber
$H\equiv\sqrt{2}\mathfrak{h}|k|a$ is roughly the disk's vertical
thickness in wavelength units. The function $A_H(z)$ is also plotted in Fig.\ \ref{AB}. 
We will be interested in a disk whose vertical thickness is
small compared to the wavelength, so $H\ll1$, and
$A_H(z)\simeq\sin(z)/z-\mbox{Ci}(z)$, where $\mbox{Ci}(z)$ is the cosine integral
of \citet{AS72}. Also keep in mind that these results assumed that the
wavenumber $k$ varies little over a single wavelength; subsection \ref{wavenumber}
will note when this approximation breaks down.

\subsubsection{ring--satellite evolution}

The ring at semimajor axis $a$ is also being perturbed by the satellite, and that
ring's disturbing function $R_s$ due to the satellite is Eqn.\ (\ref{deltaR0}) 
with $\delta m'$ replaced by the satellite's mass $m_s$:
\begin{equation}
  \label{Rs}
  R_s=-\frac{1}{4}\mu_s(na)^2\alpha\tilde{b}^{(1)}_{3/2}(\alpha)
    \left[\frac{1}{2}I^2-II_s\cos(\Omega-\Omega_s)\right],
\end{equation}
where $\mu_s=m_s/M$ is the satellite's mass in units of the central planet's mass, and
$\alpha=a_s/a=(1+\Delta+x)^{-1}\simeq1-(\Delta+x)$. The 
satellite's perturbation thus causes the ring's inclination to vary at the rate
\begin{equation}
  \left.\dot{I}\right|_{sat}=-\frac{1}{na^2I}\frac{\partial R_s}{\partial\Omega}
    =\frac{1}{4}\mu_s n\alpha \tilde{b}^{(1)}_{3/2}(\alpha)I_s\sin(\Omega-\Omega_s).
\end{equation}
For a ring in the wave--excitation zone, {\it i.e.}, near the satellite, 
\begin{equation}
  \alpha\tilde{b}^{(1)}_{3/2}(\alpha)\simeq\frac{2}{\pi(x+\Delta)^2},
\end{equation}
since the ring lies a fractional distance $x+\Delta$ away from the satellite
(see Fig.\ \ref{geometry}), with both presumably well separated
such that $\Delta\gg\mathfrak{h}$. We also write
the above longitude difference as $\Omega-\Omega_s\simeq-kax+\phi_o$,
where the angle $\phi_o$ allows for the possibility that the annulus nearest the
satellite at $x=0$ may have a longitude of ascending node that differs
from the satellite's node $\Omega_s$ by angle $\phi_o$. Thus
\begin{equation}
  \label{Idot_sat}
  \left.\dot{I}\right|_{sat}\simeq
    \frac{\mu_sI_sn}{2\pi(x+\Delta)^2}\sin(-kax+\phi_o)
\end{equation}
is the rate at which the satellite alters a ring's inclination.

\subsubsection{wave amplitude}

When the wave is in steady--state, the two $I$--excitation rates,
Eqns.\ (\ref{Idot_disk}) and (\ref{Idot_sat}), are balanced,
which yields the amplitude of the bending wave:
\begin{equation}
  \label{Iexact}
  \frac{I(z)}{I_s}=\frac{|k|a}{2}\frac{\mu_s}{\mu_d}
    \frac{\sin(z-s_k\phi_o)}{(z+|k|a\Delta)^2A_H(z)}
\end{equation}
where $s_k=\mbox{sgn}(k)$ 
and $z=|k|ax$ is the downstream distance in units of
$2\pi$ wavelengths. Far downstream, where $z\gg1$,
we expect $I(z)\rightarrow$ constant. For $z\gg1$, 
$\mbox{Ci}(z)\simeq\sin(z)/z-\cos(z)/z^2+\cal{O}$$(z^{-3})$ \citep{AS72},
so $A_H(z)\simeq\cos(z)/z^2$ downstream; see Fig.\ \ref{AB}. So if $I(z)$ is
to be a finite constant, then the longitude offset must be $\phi_o=\pm\pi/2$,
and $I/I_s\simeq -(|k|a/2)(\mu_s/\mu_d)s_k\sin\phi_o$. Of course, these inclinations
must also be positive, so 
$\sin\phi_o=\pm1=-s_k$, and the bending wave amplitude becomes
\begin{equation}
  \label{Idownstream}
  \frac{I}{I_s}\simeq\frac{|k_0|a\mu_s}{2\mu_d},
\end{equation}
where $|k_0|$ is the initial wavenumber at $x=0$, where the wave
is excited at the disk's inner edge.
To make further use of this result, we still need
the initial wavenumber $k_0$, which we get from the waves' dispersion relation.

\subsection{dispersion relation}
\label{DR}

The waves' dispersion relation is obtained from Eqn.\ (\ref{omega}), with
each term in that equation assessed below. The first
term, $ \left.\dot{\Omega}\right|_{disk}$, is the rate
at which the disk drives its own precession. Again we calculate that rate by treating
the disk as numerous narrow annuli.
The rate that annulus $a$ precesses due to the
secular perturbations from the annulus at $a'$ is
\begin{equation}
  \label{delta_dot_Omega}
  \delta\dot{\Omega}=\frac{1}{na^2I}\frac{\partial(\delta R)}{\partial I}
    =-\frac{1}{2}\mu_d'n\tilde{b}^{(1)}_{3/2}(x')
    \left[1-\frac{I'(x')}{I}\cos(\Omega-\Omega')\right]dx'
\end{equation}
where $\delta R$ is Eqn.\ (\ref{deltaR}). The total precession rate due to the disk's 
self--gravity is
$\left.\dot{\Omega}\right|_{disk}=\int_{disk}\delta \dot{\Omega}$,
where the integration proceeds across the entire disk. Again, the integrand
is a steep function of $x'$, due to the  softened Laplace coefficient,
Eqn.\ (\ref{b_approx}), which allows us to replace
the quantities $I'(x')$ and $\mu_d'(x')$ with the constants $I$ and $\mu_d$.
The disk's precession rate due to its self--gravity then becomes
\begin{eqnarray}
  \label{disk_precess0}
  \left.\dot{\Omega}\right|_{disk}&\simeq&-\frac{2}{\pi}\mu_dn\int^{\infty}_{-x}
    \frac{\sin^2(|k|ax'/2)}{x'^2+2\mathfrak{h}^2}dx'\\
  \label{disk_precess}
  &=&-B_H(|k|ax)|k|a\mu_dn
\end{eqnarray}
where
\begin{equation}
  \label{Bexact}
  B_H(z)=\frac{2}{\pi}\int_{-z}^\infty\frac{\sin^2(y/2)}{y^2+H^2}dy.
\end{equation}
The function $B_H(z)$
is a dimensionless measure of the rate at which the disk drives its own precession.
When the disk is much thinner than the wavelength, $H=\sqrt{2}\mathfrak{h}|k|a\ll1$,
and 
\begin{equation}
  B_H(z)\simeq\frac{1}{2}+\frac{1}{\pi}\mbox{Si}(z)+\frac{\cos z-1}{\pi z}
\end{equation}
where $\mbox{Si}(z)$ is the sine integral of \citet{AS72}.  
Far downstream, where $z\rightarrow\infty$, the $B_H$  integral evaluates to
\begin{equation}
  \label{B_infinity}
  B_H^{\infty}\equiv \lim_{z\rightarrow\infty}B_H(z)=\frac{1}{H}(1-e^{-H}).
\end{equation}
Note that $B_H^{\infty}$ is maximal when the dimensionless wavenumber 
is small, {\it i.e.}, $H\ll1$, for which
$B_H^{\infty}\simeq1$. But if the disk is thick,  $H\gg1$
and $B_H^{\infty}\simeq H^{-1}$, which indicates that the disk's ability to
sustain a bending wave is weakened when the disk is too thick. 

\begin{figure}
\epsscale{1.0}
\plotone{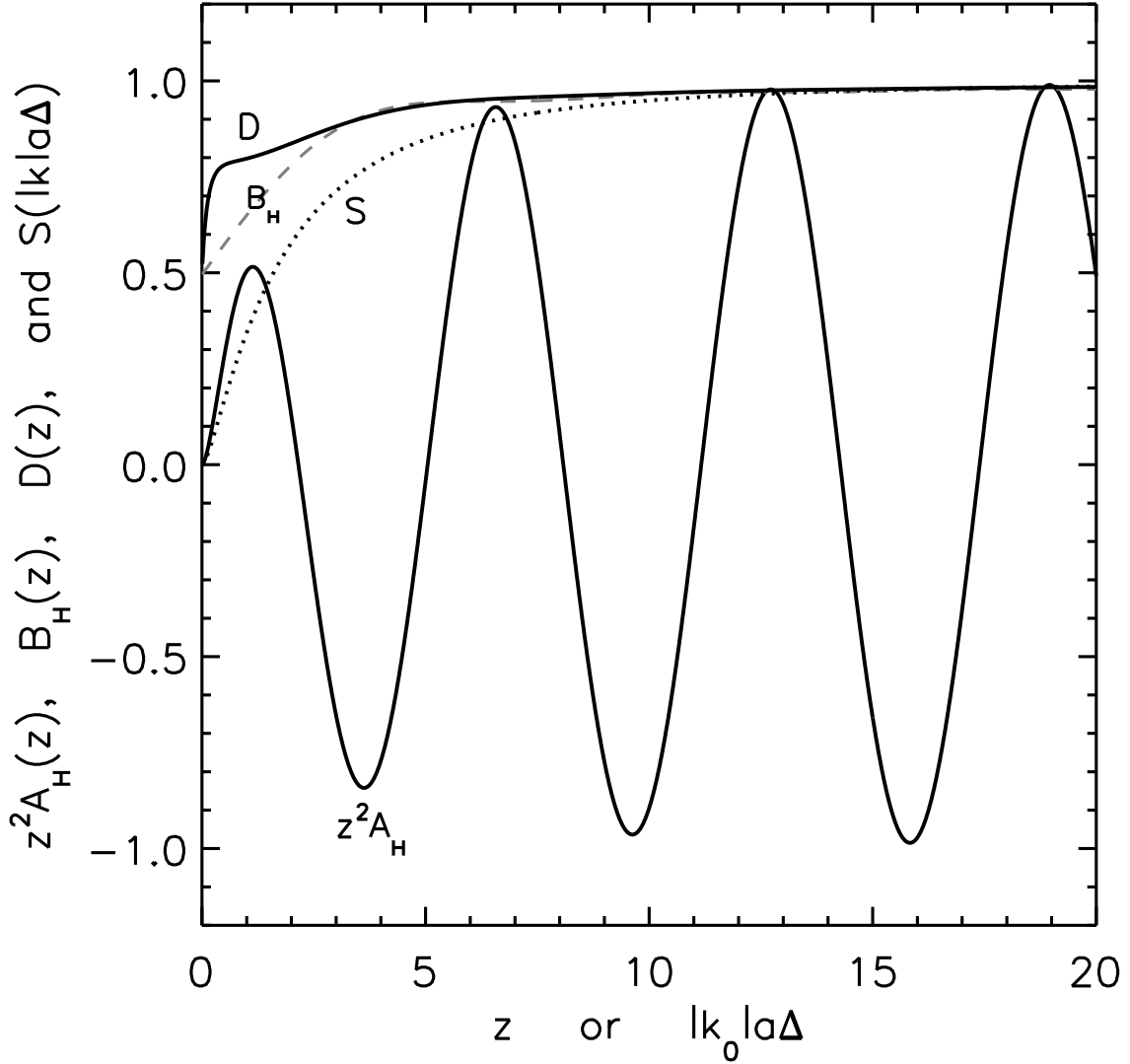}
\figcaption{
  \label{AB}
  The functions $z^2A_H(z)$ [from Eqn.\ (\ref{A}), solid curve], 
  $B_H(z)$ [Eqn.\ (\ref{Bexact}), dashed curve], 
  $D(z)$ [Eqn.\ (\ref{D}), solid curve], and
  $S(|k_0|a\Delta)$ [Eqn.\ (\ref{S}), dotted]
  are evaluated numerically for a thin disk having $H=0.01$.
  These curves are plotted versus $z=|k|ax$, which
  is the dimensionless distance from the ring's inner edge in 
  units of $2\pi$ wavelengths, or versus the dimensionless
  wavenumber $|k_0|a\Delta$.  The $D(z)$ function is evaluated with $|k|a\Delta=0.37$,
  a value that is justified in Section \ref{wavenumber}. The average of $D(z)$
  over the first wavelength, $0\le z\le2\pi$, is $\bar{D}\simeq0.87$.
  Note also that $z^2A_H(z)\simeq\cos z$ after the first wavelength.
}
\end{figure}

Figure \ref{AB} also shows a numerical evaluation
of  $B_H(z)$  for a thin disk having $H=0.01$.
This Figure shows that  $B_H(z)$ takes
values of $1/2\le B_H(z)\le1$ for $z\ge0$, with $B_H(0)=1/2$
at the disk's inner edge, and that $B_H(z)\rightarrow1$ downstream
where $z\gg1$, provided the disk is thin.

The satellite is also precessing the ring material orbiting nearest it;
that precession occurs at the rate
\begin{eqnarray}
 \left.\dot{\Omega}\right|_{sat}&=&\frac{1}{na^2I}\frac{\partial R_s}{\partial I}
    =-\frac{1}{4}\mu_sn\alpha\tilde{b}^{(1)}_{3/2}(\alpha)
    \left[1-\frac{I_s}{I}\cos(\Omega-\Omega_s)\right]\\
    \label{sat_precess}
  &\simeq&-\left[\frac{\mu_s}{2\pi(x+\Delta)^2} 
    +\frac{\mu_d\sin(|k|ax)}{\pi|k|a(x+\Delta)^2}\right]n
\end{eqnarray}
where $I_s/I$ is replaced by the downstream wave amplitude,
Eqn.\ (\ref{Idownstream}).
The first term is the familiar differential precession that would occur if the disk were
massless. The second term, which is proportional to the disk mass $\mu_d$,
is the additional precession that is due to the torque that the satellite exerts
upon the disk's spiral pattern.

The central planet's oblateness is also
driving precession; the disturbing function for that perturbation is
\begin{equation}
  R_{obl}\simeq-\frac{3}{4}J_2I^2\left(\frac{R_p}{a}\right)^2(an)^2
\end{equation}
where $J_2$ is the planet's second zonal harmonic, and $R_p$ is the planet's radius
\citep{MD99}. Precession due to oblateness is then
\begin{eqnarray}
  \label{oblate_exact}
  \left.\dot{\Omega}\right|_{obl}&=&\frac{1}{na^2I}\frac{\partial R_{obl}}{\partial I}
    =-\frac{3}{2}J_2\left(\frac{R_p}{a}\right)^2n\\
  \label{oblate_approx}
  &\simeq&\left[1-\frac{7}{2}(x+\Delta)\right]\left.\dot{\Omega}_s\right|_{obl},
\end{eqnarray}
where Eqn.\ (\ref{oblate_approx}) is a Taylor expansion of 
Eqn.\ (\ref{oblate_exact}) in the small quantity $x+\Delta$, and
$\left.\dot{\Omega}_s\right|_{obl}\equiv-(3J_2/2)(R_p/a_s)^2n_s$
is the rate at which the satellite's orbit precesses due to oblateness,
where $n_s$ is the satellite's mean motion.

Summing Eqns.\ (\ref{disk_precess}), (\ref{sat_precess}), and (\ref{oblate_approx})
provides the dispersion relation for the spiral bending waves:
\begin{equation}
  \label{omega(k)}
  \omega(|k|)\simeq-D(z)\mu_d|k|an
    -\frac{\mu_sn}{2\pi(x+\Delta)^2}
    +\left[1-\frac{7}{2}(x+\Delta)\right]\left.\dot{\Omega}_s\right|_{obl},
\end{equation}
where
\begin{equation}
  \label{D}
  D(z)=B_H(z)+\frac{\sin z}{\pi(z+|k|a\Delta)^2}.
\end{equation}
All terms in Eqn.\ (\ref{omega(k)}) are negative, so the disk
precesses in a retrograde sense. 
Note that if a spiral bending wave is to persist over time, then all parts of the disk
must precess in concert. The dispersion relation, Eqn.\ (\ref{omega(k)}), thus
tells us how the wavenumber $|k(x)|$ must adjust
throughout the disk in order for the spiral bending wave to precess coherently.

The first term in the dispersion relation
is due to the disk's self--gravity. 
That term is proportional to $D(z)$, and it has two parts:
self--precession that is driven by the bending wave itself 
(the $B_H$ term $D$),
and the additional precession that is driven by the satellite's torque on the spiral wave
pattern [the latter term in Eqn.\ (\ref{D})]. The function $D(z)$ is plotted in
Fig.\ \ref{AB}, which shows that $1/2\le D(z)\le1$.

The second term in Eqn.\ (\ref{omega(k)})
is the rate at which the satellite drives differential precession in the
disk; this effect is most prominent nearest the satellite. 
The third term is the rate at which the oblate 
central planet drives differential precession, and this occurs all throughout the disk.
Differential precession can inhibit wave--action by shredding the spiral pattern. 
But inspection of the dispersion relation suggests that bending waves can propagate,
despite differential precession due to the satellite, when the satellite's mass
is sufficiently small, {\it i.e.}, when $\mu_s\ll\mu_d|k|a\Delta^2$. The dispersion relation
also tells us that the wavenumber $|k|$ must also increase with radial distance $x$
in order to compensate for the additional differential precession that is
due to the oblate central planet.

\subsubsection{group velocity}

The waves' group velocity is \citep{T69, BT87} 
\begin{equation}
  \label{cg}
  c_g=\frac{\partial\omega}{\partial k}\simeq -s_k\mu_dan
\end{equation}
upon setting $D(z)\simeq1$ downstream; 
this is the rate at which the spiral bending wave
propagates radially \citep{H03}. Since the satellite is launching 
outward--propagating waves
from the disk's inner edge, the group velocity must be positive, which
implies that $s_k=\mbox{sgn}(k)=-1$. Spiral waves having $k<0$
are called {\em leading} waves. 
Note also that $\sin\phi_o=-s_k=+1$, so $\phi_o=\pi/2$,  
which means that the longitude of ascending node at the disk's inner
edge leads the satellite's node by $90^\circ$.

\subsubsection{wavenumber $k$}
\label{wavenumber}

The wavenumber $k$ can be obtained by calculating the satellite's precession
rate $\dot{\Omega}_s$. When the system is in steady--state, both the satellite
and the spiral wave precess at the same rate, $\dot{\Omega}_s=\omega(|k|)$,
which provides another equation for the wavenumber $k$.

The satellite's node $\Omega_s$ is being 
precessed by the disk and by the central planet, so
\begin{equation}
  \dot{\Omega}_s=\left.\dot{\Omega}_s\right|_{disk}
    + \left.\dot{\Omega}_s\right|_{obl}
\end{equation}
where $\left.\dot{\Omega}_s\right|_{disk}=\int_{disk}\delta\dot{\Omega}_s$
is the satellite's precession rate due to the entire disk, and where
\begin{equation}
  \delta\dot{\Omega}_s=-\frac{1}{2}\mu_d'n_s\tilde{b}^{(1)}_{3/2}(\Delta+x')
    \left[1-\frac{I'}{I_s}\cos(\Omega_s-\Omega')\right]dx'
\end{equation}
is the satellite's precession rate due to a disk annulus of radius $a'$ and mass
$\delta m'$. This can be obtained from Eqn.\ (\ref{delta_dot_Omega}) with
$n, a, I, \Omega$ replaced by $n_s, a_s, I_s, \Omega_s$ and 
the separation $x'\rightarrow x'+\Delta$. The satellite's precession rate
due to the entire disk is
\begin{mathletters}
\begin{eqnarray}
  \left.\dot{\Omega}_s\right|_{disk}&=&\int_{disk}\delta\dot{\Omega}_s\simeq
    -\frac{1}{\pi}\mu_dn_s\int_0^\infty(x'+\Delta)^{-2}
    \left[1-\frac{I}{I_s}\cos(kax'-\phi_o)\right]dx'\\
  \label{dot_Omega_s0}
  &\simeq&-\frac{\mu_dn_s}{\pi\Delta}-
    \frac{\mu_sn_s}{2\pi\Delta^2}S(|k_0|a\Delta)
\end{eqnarray}
\end{mathletters}
where
\begin{equation}
  \label{S}
  S(|k_0|a\Delta)\equiv |k_0a\Delta|^2\int_0^\infty\frac{\sin(y)dy}{(y+|k_0|a\Delta)^2}.
\end{equation}
The first term in Eqn.\ (\ref{dot_Omega_s0}) is the rate at which the undisturbed
disk precesses the satellite's orbit. The second term is the rate at which
the bending wave, whose amplitude is proportional to 
$\mu_s$ by Eqn.\ (\ref{Idownstream}), drives additional precession. 
The $S$ function in that term is a dimensionless measure
of the wave's contribution to the satellite's precession rate; that quantity depends on
the wave's initial wavenumber $|k_0|$, and is plotted in 
Fig.\ \ref{AB}, which shows that $0\le S(|k_0|a\Delta)\le 1$.

Note that if the satellite instead orbited at the center of a narrow gap in the disk,
then the first term
in Eqn.\ (\ref{dot_Omega_s0}) would be doubled due to the disk matter orbiting
interior to the satellite. We might also expect additional precession to occur
due to any bending waves launched in this interior disk, 
but it will be shown below that
this contribution is unimportant. With this in mind, 
we will generalize Eqn.\  (\ref{dot_Omega_s0})
to account for a possible inner disk by writing
\begin{equation}
    \label{dot_Omega_s}
  \left.\dot{\Omega}_s\right|_{disk}\simeq
    -\frac{\varepsilon\mu_dn_s}{\pi\Delta}
    -\frac{\mu_sn_s}{2\pi\Delta^2}S(|k_0|a\Delta)
\end{equation}
where it is understood that $\varepsilon=1$ if the disk is entirely exterior to the
satellite, and that $\varepsilon=2$ if the satellite instead orbits in the center
of a gap whose fractional half--width is $\Delta$. The satellite's total
precession rate then becomes
\begin{equation}
  \dot{\Omega}_s= \left.\dot{\Omega}_s\right|_{disk}+
     \left.\dot{\Omega}_s\right|_{obl}=
    -\frac{\varepsilon\mu_dn_s}{\pi\Delta}-\frac{\mu_sn_s}{2\pi\Delta^2}
    S(|k_0|a\Delta)+ \left.\dot{\Omega}_s\right|_{obl}.
\end{equation}

When the disk and satellite are in steady--state, the satellite and its spiral bending
pattern precess in concert, so $\dot{\Omega}_s=\omega(|k|)$,
which after some manipulation yields the dispersion relation
\begin{equation}
  \label{DR0}
  \pi D(z)|k|a\Delta=\varepsilon+
    \frac{\mu_c}{\mu_d}\left(1+\frac{x}{\Delta}\right)
    +\frac{\mu_s}{2\mu_d\Delta}f(|k_0a\Delta|, z),
\end{equation}
where
\begin{equation}
  f(|k_0a\Delta|, z)=S(|k_0|a\Delta)-\frac{|k_0a\Delta|^2}{(|k_0a\Delta|+z)^2}
\end{equation}
is another function of distance $z$ and wavenumber $|k_0|$,
one that is restricted to the interval $-1\le f\le 1$, and
\begin{equation}
  \label{mu_c}
  \mu_c\equiv\frac{21\pi}{4}
    \left(\frac{R_p\Delta}{a_s}\right)^2J_2,
\end{equation}
which will be called the critical disk mass.

\subsubsection{limits on wave propagation}
\label{limits}

The disk's ability to sustain these bending waves is assessed by
multiplying the dispersion relation (\ref{DR0})
by $\sqrt{2}\mathfrak{h}/\pi\Delta$, which yields
\begin{equation}
  \label{DR1}
  HD=\frac{\sqrt{2}\mathfrak{h}}{\pi\Delta}\left[
    \varepsilon+\frac{\mu_c}{\mu_d}\left(1+\frac{x}{\Delta}\right)
    +\frac{\mu_sf}{2\mu_d\Delta}\right]
\end{equation}
where the dimensionless wavenumber
$H=\sqrt{2}\mathfrak{h}|k|a$. Far downstream, where $z\gg1$
and $D(z)= B_H^\infty=(1-e^{-H})/H$ 
(see Eqns.\ \ref{B_infinity} and \ref{D}), so $HD=1-e^{-H}<1$.
Wave propagation thus requires the
right-hand side of Eqn.\ (\ref{DR1}) to always be less than unity,
which places an upper limit on the thickness of a disk that is able to sustain
these bending waves, namely, that 
$\mathfrak{h}< \mathfrak{h}_{\mbox{\scriptsize max}}$ where
\begin{equation}
  \label{max_h}
  \mathfrak{h}_{\mbox{\scriptsize max}}\equiv\frac{\pi\Delta/\sqrt{2}}
    {\varepsilon+\mu_c/\mu_d+\mu_s/2\mu_d\Delta},
\end{equation}
upon setting $x=0$ and $f=1$ in order to obtain the most
conservative limit on the disk's fractional thickness 
$\mathfrak{h}_{\mbox{\scriptsize max}}$.

The remainder of this paper
will assume that the disk is thin enough to sustain
density waves, namely, that 
$\mathfrak{h}\ll\mathfrak{h}_{\mbox{\scriptsize max}}$, or
equivalently that $H\ll1$, so that $D\sim{\cal O}(1)$. Also 
recall that Section \ref{DR}
anticipated a wave solution to occur when the satellite's mass is small. Specifically,
when $\mu_s\ll2\varepsilon\mu_d\Delta$, the right-most terms in Eqns.\ (\ref{DR0})
or (\ref{DR1}) may be neglected, which then provides the wavenumber $k$
as a simple function of distance $x$ in the disk:
\begin{equation}
  \label{k}
  |k|\simeq\frac{1}{\pi\bar{D}a\Delta}
    \left[\varepsilon+\frac{\mu_c}{\mu_d}\left(1+\frac{x}{\Delta}\right)\right]
\end{equation}
where $D(z)$ has been replaced with its average value over the first 
wavelength, $\bar{D}$. And if the disk is sufficiently massive, namely,
that $\mu_d\gtrsim18\mu_c/\varepsilon$, then
the initial wavenumber at $x=0$ is 
$|k_0|a\Delta\simeq\varepsilon/\pi \bar{D}\simeq0.37$, 
where $\varepsilon=1$ and
$\bar{D}\simeq0.87$ according to Fig.\ \ref{AB}. 
In that limit, the first wavelength
is $\lambda_0=2\pi/|k_0|\simeq2\pi^2\bar{D}\Delta a\simeq17\Delta\cdot a$.
However shorter wavelengths will result when $\mu_d$
does not exceeds the above threshold.

Plugging Eqn.\ (\ref{k}) evaluated at $x=0$
into Eqn.\ (\ref{Idownstream}) then yields the wave amplitude
in terms of the system's physical parameters:
\begin{equation}
  \label{I}
  \frac{I(x)}{I_s}\simeq
    \frac{\mu_s(\varepsilon+\mu_c/\mu_d)}{2\pi\bar{D}\mu_d \Delta}.
\end{equation}

\subsubsection{inclination damping}
\label{Idamping}

The satellite launches a spiral bending wave via
its secular gravitational perturbations of the ring.
Those perturbations tilt the orbital plane of the nearby ring particles, and they in turn
tilt the orbits of the more distant parts of the disk. 
Tilting an annulus in the disk also tips that ring's angular momentum
vector, so the excitation of a bending
waves transmits {\em in--plane} angular momentum 
from the satellite to the disk. Consequently,
wave--excitation damps the satellite's inclination $I_s$, and that rate can be 
calculated using the Lagrange planetary equation for $\dot{I}_s$.

\begin{figure}
\epsscale{1.0}
\plotone{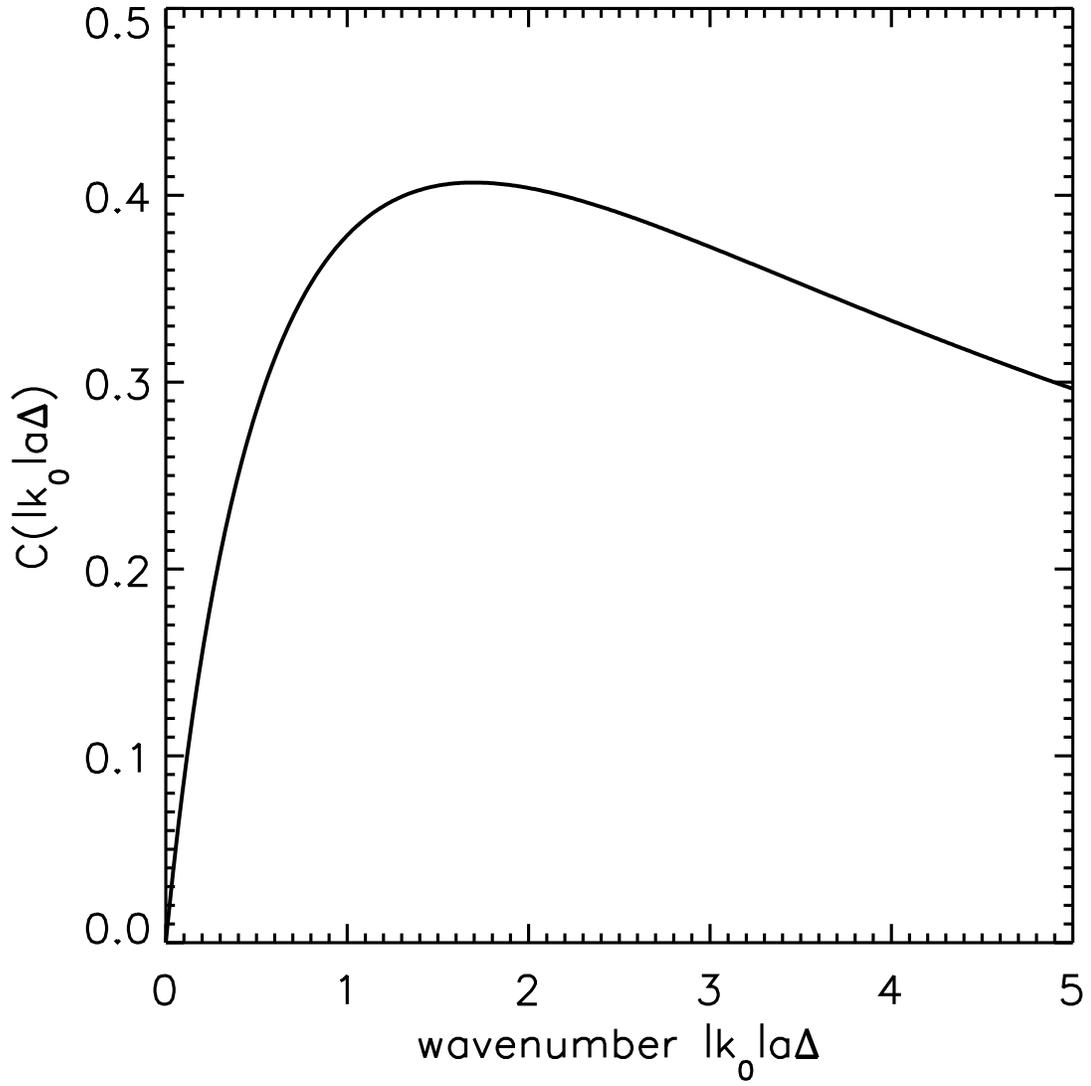}
\figcaption{
  \label{Cfigure}
  The function $C(|k_0|a\Delta)$ from Eqn.\ (\ref{C}),  
  plotted versus the dimensionless
  wavenumber $|k_0|a\Delta$.
}
\end{figure}

The rate $\delta\dot{I}_s$ at which a single annulus in the disk damps 
the satellite's inclination $I_s$ is Eqn.\ (\ref{delta-dot-I}), again with 
$n, a, I, \Omega\rightarrow n_s, a_s, I_s, \Omega_s$ and $x'\rightarrow x'+\Delta$.
Integrating the contributions by all annuli in the disk gives the satellite's
total inclination--damping rate,
\begin{equation}
  \label{Idamping0}
  \dot{I}_s=\int_{disk}\delta\dot{I}_s
    =\frac{\mu_dIn_s}{\pi k_0a\Delta^2}C(|k_0|a\Delta)
\end{equation}
where $k_0$ is the initial wavenumber at the disk's inner edge, and the function
\begin{equation}
  \label{C}
  C(z)=z^2\int_0^\infty
    \frac{\cos(y)dy}{(y+z)^2}
    =z+z^2\left\{\left[\mbox{Si}(z)-\frac{\pi}{2}\right]\cos z-\mbox{Ci}(z)\sin z\right\}
\end{equation}
is shown in Fig.\ \ref{Cfigure}. Inserting Eqn.\ (\ref{Idownstream}) into
(\ref{Idamping0}) and noting that $k_0=-|k_0|$
then provides the inclination--damping rate in terms
of the system's physical parameters:
\begin{equation}
  \label{Idamping1}
  \frac{\dot{I}_s}{I_s}=-\frac{C(|k_0|a\Delta)}{2\pi}\frac{\mu_s}{\Delta^2}n_s.
\end{equation}
The reciprocal of the above gives the e--fold timescale for the satellite's
inclination decay:
\begin{equation}
  \label{tau_i}
  \tau_i=\frac{\Delta^2P_{orb}}{C(|k_0|a\Delta)\mu_s},
\end{equation}
where $P_{orb}=2\pi/n_s$ is the satellite's orbit period.
Note that for a planet that is not too oblate (e.g., Section \ref{wavenumber})
$|k_0|a\Delta=0.37$, so $C(|k_0|a\Delta)\simeq0.24$ (Fig.\ \ref{Cfigure}).
This inclination--damping rate is also confirmed below, in a numerical simulation
of spiral bending waves launched in a planetary ring.

\section{Simulations of spiral bending waves}
\label{rings}

The rings model of \citet{H03} will be used to confirm the preceding results.
The rings model treats the system as a set of N discrete gravitating annuli having
semimajor axes $a_j$, inclinations $I_j$,  nodes $\Omega_j$, and
half--thicknesses $h_j$. The model only considers the system's secular 
gravitational perturbations, so it also solves the same equations of motion,
Eqns. (\ref{EOM}), but the model does so without making any of 
the wave--assumptions invoked in Section \ref{amplitude}.
The model thus provides an independent check of the analytic results obtained above. 

\subsection{waves in an exterior disk}
\label{exterior}

The rings model is used to simulate the spiral bending waves that are
launched by an inclined satellite that orbits just interior to a disk.
Figure \ref{Iwave} shows the amplitude of this bending wave 
as it advances across a disk. The system's parameters are detailed in
Fig. \ref{Iwave}. Those parameters do not correspond to any real ring--satellite
system; rather, these parameters were chosen to illustrate the results of
Section \ref{EOM} in the limit in which those results
were obtained, namely, that the satellite's mass is small, {\it i.e.}, 
$\mu_s\ll2\mu_d\Delta$ so that Eqn.\ (\ref{k}) is valid.
Those parameters were also chosen so that the factor $\mu_c/\mu_d$
appearing in the wavenumber Eqn.\ (\ref{k}) is 0.2, 
which causes the wavelength to slowly decrease with distance $x$
as they propagate away.
Nonetheless, the simulation reported in Fig. \ref{Iwave} does correspond
loosely to a small $\sim10$ km satellite orbiting just interior 
to ring whose surface density is similar to Saturn's main A ring.

Inspecting this system's angular momentum provides a quick check on the quality
of this calculation. This system should conserve the
in--plane component of its total angular momentum, 
$L_i=\onehalf\sum m_j n_j a_j^2 I_j^2$, 
where the sum runs over all rings and satellites in the
system \citep{H03}. The single--precision calculation shown in Fig.\ \ref{Iwave}
conserves $L_i$ with a fractional error of $|\Delta L_i/L_i|<2\times10^{-5}$.

Note that the time for these waves to propagate a 
fractional radial distance $x=\Delta r/a$ is
\begin{equation}
  t_{prop}=\frac{\Delta r}{c_g}=\frac{xP_{orb}}{2\pi\mu_d},
\end{equation}
where $c_g$ is the waves' group velocity, Eqn.\ (\ref{cg}).
The simulated disk has a normalized mass of $\mu_d=5\times10^{-8}$
and a fractional width $x=0.02$, so the anticipated propagation time
is $t_{prop}=64\times10^3$ orbits, which compares favorably with the
simulation (see Fig. \ref{Iwave}).

If the disturbance seen in Fig.\ \ref{Iwave} is indeed a spiral bending wave,
then the disk's longitude of ascending node $\Omega(a)$ should steadily advance
as $a$ increases across the disk.  This is confirmed in Fig. \ref{wk}, which 
shows the waves' longitudes relative to the satellite's,
$\Omega(a)-\Omega_s$. These are also the longitudes where the 
warped disk passes through the central planet's equatorial plane. 
Note that this disk will have its maximum
elevation at longitudes $90^\circ$
ahead of that seen in Fig. \ref{wk}, with its minimum elevation
at longitudes $90^\circ$ behind. Note also that the longitude
of the disk's inner edge is $90^\circ$ ahead of the satellite's longitude, as
expected. And since the wavenumber $k=-\partial\Omega/\partial a$ is
negative, this spiral pattern is indeed a leading wave.
We also note that once the bending wave is established in the
disk, the disk's longitudes precess at the same rate as the satellite's,
{\it i.e.}, $\dot{\Omega}(a)=\dot{\Omega}_s$, and that the disk's
inclinations are constant, $\dot{I}(a)=0$, which justifies our 
steady--state assumptions, Eqn.\ (\ref{ss}).

Figure \ref{wk} also plots the dimensionless wavenumber $|k|a\Delta$ across
the disk at time $t=75\times10^{3}$. This is the moment when the bending
wave is just starting to reflect at the disk's outer edge, which accounts for the curve's
raggedness there. Also plotted
is the expected wavenumber, Eqn.\ (\ref{k}), which compares favorably.

The rate at which the disk damps the simulated satellite's inclination $I_s$ is shown
in Fig.\ \ref{idot}, where it is compared to the expected rate, Eqn.\ (\ref{Idamping1}). 
That rate is calculated by noting that the waves' 
initial wavenumber is $|k_0|a\Delta\simeq0.63$ at the disk's inner edge
(see Fig.\ \ref{wk}), so the $C$
that appears in Eqn.\ (\ref{Idamping1})
is $C(|k_0|a\Delta)=0.32$, according to Fig.\ \ref{Cfigure}. The expected and observed
inclination damping rates are in good agreement.

\subsection{Satellite in a gap}
\label{gap}

The simulation described by Figs.\  \ref{Iwave} and \ref{wk}
is a bit of fiction, since there are no known
satellites orbiting just interior to a broad planetary
ring. For instance, all of the major Saturnian 
satellites orbit exterior to Saturn's main rings.
However there are two noteworthy exceptions: 
the small satellite Pan, which orbits in the Encke gap
in Saturn's A ring, and Daphnis, 
which inhabits the Keeler gap in Saturn's A ring \citep{P05}.

A simulation of an inclined Pan as it orbits in the 
Encke gap is reported in Figure \ref{pan1}, which shows the state of this system
at time $t=7.5\times10^4$ orbits. This is the time required for Pan to launch
a leading spiral bending wave at the gap's outer edge that then propagates
to the simulated ring--system's outer edge.
Figure \ref{pan1} shows that waves' initial wavelength is 
$\lambda_0=0.0037a_s\simeq500$ km, and that their wavelength shrinks
with distance $x$ as they propagate towards the outer edge of the A ring,
which lies a fractional distance $x=0.024$ away.
So if these waves are not otherwise 
damped en route by collisions among ring particles,
their wavenumber will have grown to $|k|a\simeq1.1\times10^4$
when they reach the A ring's outer edge (see Eqn.\ \ref{k}), which
corresponds to wavelength of 
$\lambda=2\pi/|k|\simeq80$ km.

Figure \ref{pan2} also shows the system at the
later time $t=1.5\times10^5$ orbits, which is when the
wave has since reflect at the simulated ring's outer edge and returned to the
launch site. Here we see the superposition of an 
outbound leading wave with an inbound 
trailing wave, which results in a standing bending wave
throughout the disk. As the Figure shows, when the standing wave emerges, it arranges
the disk's longitudes $\Omega$ such that they
alternate between $-90^\circ$ and  $+90^\circ$ of the satellite's node
$\Omega_s$ at every half--wavelength. So 
if the waves launched by an inclined Pan do not get damped downstream, 
that bending wave will reflect at the outer A ring and
return to Pan's vicinity where it can communicate its in--plane angular
momentum back to the satellite. At this moment, inclination--damping then ceases.

Figures \ref{pan1} and \ref{pan2} also show  that Pan does not launch any 
inward--propagating
waves. Although the satellite's secular perturbations do excite inclinations 
at the gap's inner edge, those
disturbances do not travel further inwards.
The wavenumber $|k|$ for any disturbance that might propagate in
 the interior disk is\footnote{This wavenumber 
is obtained by repeating the derivation of Section
\ref{EOM} by applying that to an annulus in the inner disk whose semimajor
axis is again $a=a_s(1+x+\Delta)$, but with the distances
$x$ and $\Delta$ now understood as
having negative values. 
So when integrating the net perturbation that the entire disk exerts
on an annulus, the integration variable $x'$ in 
Eqns.\ (\ref{Idot_disk0}) and (\ref{disk_precess0})
now runs over $-\infty\le x' \le -x$.
The net effect of this is to merely change the sign on certain terms:
$\dot{I}_{disk}$ is $-1\times$Eqn\ (\ref{Idot_disk}), 
and $s_k=\mbox{sgn}(k)=\sin\phi_i=+1$,
where $\phi_i=\pi/2$ is the longitude offset between the inner gap edge and the satellite.
Any inward--propagating waves in the inner disk are trailing, since $k>0$.
The argument of the sinusoid in Eqn.\ (\ref{sat_precess}) also changes sign.
Accounting for these sign changes then yields  Eqn.\ (\ref{k_in}).}
\begin{equation}
  \label{k_in}
  |k|\simeq\frac{1}{\pi\bar{D}a|\Delta|}
    \left[\varepsilon-\frac{\mu_c}{\mu_d}\left(1+\left|\frac{x}{\Delta}\right|\right)\right],
\end{equation}
which is identical to Eqn.\ (\ref{k}) except for the sign on the disk mass term. 
Since the right hand side must be positive,
Eqn.\ (\ref{k_in}) tells us that waves in the inner disk can only propagate
in the zone where $|x|<x_{in}$, where
\begin{equation}
  \label{x_in}
  x_{in}\equiv\left(\frac{\varepsilon \mu_d}{\mu_c}-1\right)|\Delta|
\end{equation}
is the distance of the waves' maximum excursion inwards of the satellite's orbit.
Getting waves to propagate inwards a significant distance thus requires
the disk mass to be sufficiently high, namely, $\mu_d\gg\mu_c/2$,
where $\varepsilon=2$ for a gap--embedded satellite. 
Saturn's A ring has a disk mass of $\mu_d\sim5\times10^{-8}$
and a critical disk mass of $\mu_c=7.8\times10^{-8}$
(from Fig.\ \ref{pan1} caption), so $\mu_d\gg\mu_c/2$ is not well--satisfied, and
inward--propagating bending waves are precluded. 

Also note that this simulation does not satisfy $\mu_d\gtrsim18\mu_c/\varepsilon$,
which means that the wavenumber $k$ does varies 
substantially across that first wavelength (see Eqn.\ \ref{k}). 
A wavenumber that is nearly constant over that first wavelength
is of course a key assumption of Sections
\ref{EOM}--\ref{rings}, so the analytic results obtained there might seem
not apply to Pan.
Nonetheless, when those formulas are compared to the model results, we find that
Eqn.\ (\ref{k}) to be in excellent agreement with the wavenumber $k(x)$ exhibited
by the simulated wave. But  Eqn.\ (\ref{I}) does overestimate
the amplitude of this simulated wave by a factor of $\sim4$, which in turn causes
Eqn.\ (\ref{Idamping1}) to overestimate Pan's inclination--damping rate by the same
factor.

However Daphnis inhabits the 
narrower Keeler gap, whose fractional half width $\Delta=1.1\times10^{-4}$
is a tenth that of the nearby Encke gap, so its
$\mu_c$ is 100 times smaller, and $x_{in}\simeq120|\Delta|=0.013$,
which corresponds to a physical distance of $x_{in}a_s\simeq1800$ km,
or about 10 wavelengths. So when
the rings model is used to simulate the spiral waves that an inclined Daphnis
would launch, we do indeed see a wave launched from the gap's inner edge. That wave
propagates inwards approximately a distance $x_{in}$, where it reflects and propagates
outwards and across the Keeler gap. That satellite's simulated 
$I$--damping timescale is also in 
good agreement with the prediction, Eqn.\ (\ref{tau_i}).
 
\section{External Vertical Resonances}
\label{vertical}

A satellite orbiting near a planetary ring also excites inclinations at its many
external vertical resonances in the ring. This also communicates in--plane
angular momentum between the satellite and the ring, but in a manner that
{\em excites} the satellite's inclination $I_s$. \cite{BGT84} calculate the rate
at which the external resonances in a narrow ring of mass $\delta m'$
excite the satellite's inclination:
\begin{equation}
  \delta \dot{I}_s=g\mu_sn_s |x'|^{-5}\frac{\delta m'}{M}I_s
\end{equation}
where $g=0.0118$, and $x'$ is the satellite's fractional distance from the ring of mass
$\delta m'=2\pi\sigma a^2 dx'$.
The satellite's total $I$--excitation rate is the above integrated
across the entire disk, $\dot{I}_s=\int_{disk}\delta \dot{I}_s$.
If the satellite orbits in the center of a gap in a broad planetary ring, 
the total excitation rate due to external vertical resonances is
\begin{equation}
  \frac{\dot{I}_s}{I_s}=\frac{g\mu_s\mu_d n_s}{\Delta^4}
\end{equation}
\citep{WH03}.
So if the satellite's orbit
is to remain confined to the ring plane, this $I$--excitation due to
the satellite's external resonances must be smaller than the $I$--damping
that results from its secular interaction with the ring, Eqn.\ (\ref{Idamping1}).
Comparing these two rates shows that the satellite's inclination is stable, {\it i.e.},
$\dot{I}_s<0$, when its gap is sufficiently large, namely, when
\begin{equation}
  \label{gap_width}
  \Delta^2>\frac{2\pi g\mu_d}{C(|k_0|a\Delta)}.
\end{equation}
Bending waves launched by Pan and Daphnis have initial wavenumbers
of $|k_0|a\Delta\simeq1$ (see Eqn.\ \ref{k}), 
so $C(|k_0|a\Delta)\simeq0.3$ (see Fig.\ \ref{Cfigure}),
and the requirement for inclination damping becomes $\Delta\gtrsim0.5\sqrt{\mu_d}$.
These satellites inhabit Saturn's A ring, which has a normalized disk mass
of $\mu_d\simeq5\times10^{-8}$, so their inclinations are stable if their
gap half--widths are wider than $\Delta\gtrsim1.1\times10^{-4}$. Pan easily
satisfies this requirement ($\Delta=0.0012$), while Daphnis marginally so
($\Delta=1.1\times10^{-4}$). The $I$--damping timescale for Pan
is quite short, only $\tau_i\simeq1.7\times10^6$ orbits 
(e.g., four times Eqn.\ \ref{tau_i}), 
which corresponds to $\tau_i\simeq2700$ years. A comparable $I$--damping
timescale is also obtained for Daphnis, whose size is about four times smaller than
Pan's \citep{Setal06}, and whose gap is ten times narrower.

\section{Summary and Conclusions}
\label{summary}

The secular perturbations exerted by an inclined satellite orbiting in a gap in a 
broad planetary ring tends to excite the inclinations of the nearby ring particles.
The ring's self gravity then allows that disturbance
to radiate away in the form of a spiral bending
wave. The wavelength $\lambda=2\pi/|k|$ of any outbound waves
is obtained from  Eqn.\ (\ref{k}), which shows that $\lambda$
decreases with distance $x$ from the nearby gap edge. 
These wavelength variations are due to a competition between
the disk's self gravity and 
the differential precession that is due to the oblate central planet. As an example,
we find that an inclined Pan, which inhabits
the Encke gap in Saturn's main A ring, would excite a bending 
wave having an initial wavelength of about 500 km. 
And if that wave manages to propagate out to the outer edge of the
A ring without damping, the wavelength will then have shrunk
down to about 80 km there.

A gap--embedded satellite will also try to launch a wave at the gap's inner edge,
but the range of these waves
is limited by how far they can propagate until their wavenumber
has shrunk to zero; see Eqn.\ (\ref{k_in}). That distance is controlled 
by the width of the gap, with a narrower gap resulting in a greater inward  
excursion; see Eqns.\ (\ref{x_in}) and  (\ref{mu_c}). 
For instance, Pan is unable to launch an
inward--propagating wave, while Daphnis, which inhabits the
narrower Keeler gap, could launch a disturbance that propagates inwards
about $1\%$ of its orbit before reflecting and propagating back out
and across the Keeler gap. 

The amplitude of this wave is given by Eqn.\ (\ref{I}),
and the excitation of this wave also damps the satellite's inclination quite vigorously,
at a rate given by Eqn.\ (\ref{Idamping1}). 
This $I$--damping mechanism also competes
with the satellite's many other vertical resonances in the ring, which try to pump up
the satellite's inclination \citep{BGT84}. However the secular
$I$--damping will dominate when
the satellite's gap is sufficiently wide, namely, when Eqn.\ (\ref{gap_width})
is satisfied. Saturn's gap--embedded moon Pan satisfies
this requirement, while Daphnis is at the threshold.
This secular phenomenon also damps
inclinations on a very short timescale $\tau_i$, which for these satellites
is of order 3000 yrs (Eqn.\ \ref{tau_i}). This of course assumes that these
waves damp somewhere downstream,
rather than reflecting at the ring's outer edge and returning to the launch site. 
But if these waves reflect and return without suffering
significant damping, then a standing
wave will emerge in the disk. That standing wave would also communicate
some of its in--plane angular momentum back into the satellite's orbit, so further
inclination--damping would cease. But if these waves instead damp downstream, then
this secular phenomenon represents an important stabilizing influence that tends to
confine a satellite's orbit to the ring plane. But this inclination--damping
also shuts off any subsequent wave generation,
so is seems unlikely that these waves might ever be observed in a planetary ring.

Finally, we note that the rings model employed here played a important
role in guiding the analytic results obtained above. 
The model itself is a fairly easy--to--use set of IDL
scripts, and other applications of this code are possible. 
For instance,
the rings model has revealed that an eccentric satellite's secular perturbations can
launch spiral density waves in a nearby ring, and  
the eccentricity damping that is associated with that phenomena will be assessed in
a followup study \citep{H07}. A copy of the rings model algorithm
will also be made available to others upon request.

\acknowledgments

\begin{center}
  {\bf Acknowledgments}
\end{center}

This research was initiated while the author was 
in residence at Saint Mary's University,
and that portion of this research was supported by a Discovery Grant
from the Natural Sciences and Engineering Research Council of Canada (NSERC).
This work was also supported by a grant from NASA's Outer Planets
Research Program.
The author also thanks Jayme Derrah for composing Figure \ref{geometry}.

\appendix
\section{Appendix \ref{appendix}}
\label{appendix}

The disk's vertical displacement is
$z(a, \theta, t)=a\sin I\sin(\theta-\Omega)$,
where $(a, \theta)$ are the radial and azimuthal coordinates
in the disk, and the inclination $I$ and ascending node $\Omega$
should be regarded as functions of distance $a$ and time $t$. 
When a spiral bending wave is present in the disk, the longitudes
have the form
\begin{equation}
  \Omega(a, t)=\Omega(a_0, 0)-\int^a_{a_0} k(r)dr+\omega t,
\end{equation}
where $\Omega(a_0, 0)$ is the longitude of the ascending node
at some reference distance $a_0$ at time $t=0$, $k(a)$ is the wavenumber
of the spiral bending wave, and $\omega$ is the angular rate at which the spiral
pattern rotates, also known as the pattern speed. The signs in the above expression
are chosen to follow the convention that a $k<0$ spiral
is a {\em leading} spiral, which means that a curve having a constant 
$z(a, \theta)$ in the disk
traces a spiral that advances in $\theta$ as $a$ increases. Thus
the wavenumber $k$ can be written as
\begin{equation} 
  \label{k_appendix}
  k=-\frac{\partial\Omega}{\partial a},
\end{equation}
and the group velocity is
\begin{equation}
  c_g=\frac{\partial\omega}{\partial k}
\end{equation}
\citep{T69, BT87}.

\bibliography{biblio}

\newpage

\begin{figure}
\epsscale{1.0}
\plotone{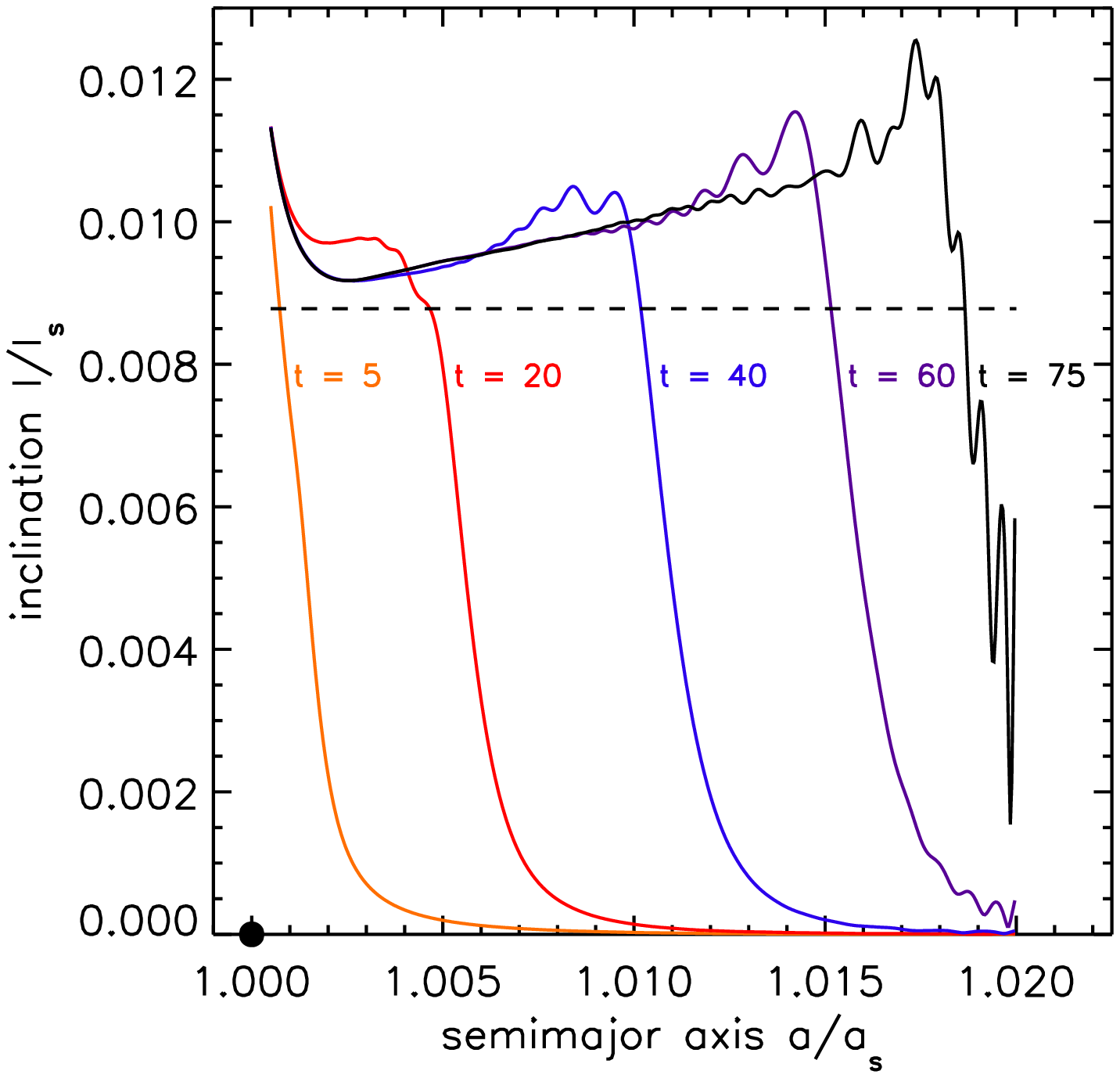}
\end{figure}

\begin{figure}
\figcaption{
  \label{Iwave}
The rings model is used to simulate bending waves
launched by an inclined satellite that orbits just interior to a disk.
The satellite's normalized mass is $\mu_s=10^{-12}$, and the disk is comprised
of $N=500$ rings having semimajor axes distributed
over $1+\Delta\le a_j/a_s\le 1.02$, where $\Delta=5\times10^{-4}$ is the fractional
distance between the satellite and the innermost ring. The rings' fractional masses are
$\mu_r=3.9\times10^{-12}$, so the normalized disk mass is
$\mu_d=\pi\sigma r^2/M=\mu_r/2(\delta/a_s)=5\times10^{-8}$,
where the rings' fractional separations are $\delta/a_s=0.02/N=4\times10^{-5}$.
The rings' fractional half--widths $\mathfrak{h}$ is also set equal to
 their separations $\delta/a_s$.
The central planet's zonal harmonic is $J_2=0.012$ and the planet's radius
is $R_p/a_s=0.45$, so the system's critical disk mass is
$\mu_c=1.0\times10^{-8}$ and $\mu_c/\mu_d=0.2$.
The satellite's initial inclination is 
$\sin I_s=10^{-5}$, with all other rings initially having
zero inclinations.  The curves show the fractional amplitude of the bending wave,
$I(a)/I_s$, as it advances across the disk, shown at selected times $t$ in units 
of $10^3$ orbital periods. The dashed line is the expected wave amplitude,
Eqn.\ (\ref{I}), with $\varepsilon=1$. 
}
\end{figure}

\begin{figure}
\epsscale{1.0}
\plotone{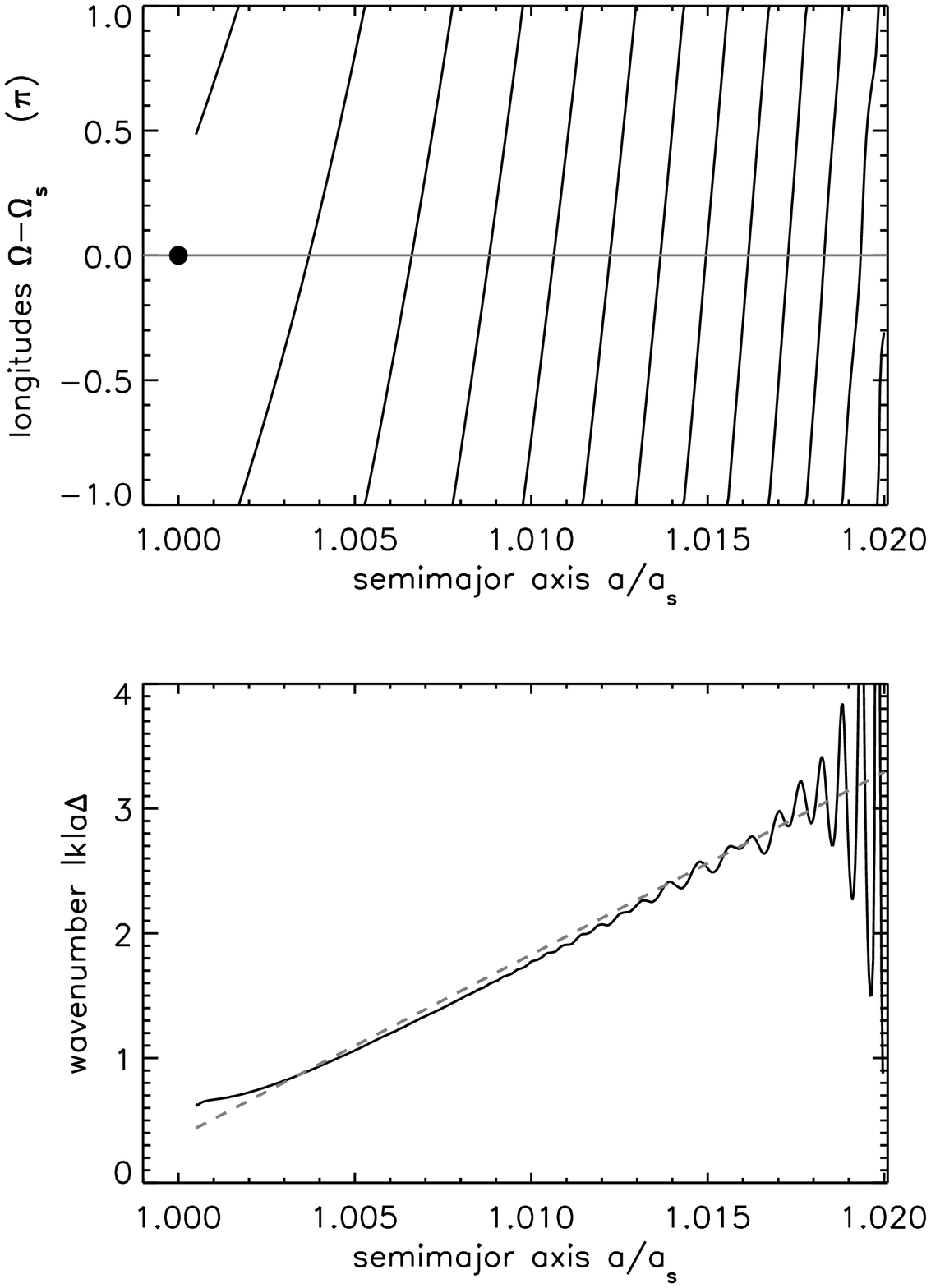}
\end{figure}

\newpage
\begin{figure}
\figcaption{
  \label{wk}
  The upper figure shows the disk's longitude of ascending nodes
  $\Omega(a)$ relative to the satellite's node $\Omega_s$, in units of $\pi$,
  for the simulation of Fig.\ \ref{Iwave} at time $t=75\times10^3$ orbits,  
  when the wave has swept across the disk.
  The lower figure shows the dimensionless wavenumber $|k|a\Delta$ 
  at this moment, where wavenumber is calculated from
  $k=-\partial\Omega/\partial a$. Note that the simulated curve gets a bit ragged
  at the disk's outer edge, which is where the bending wave is just starting to
  reflect. The dashed line is the expected wavenumber,
  Eqn.\ (\ref{k}), with $\varepsilon=1$. This spiral pattern has an initial wavenumber
  of about $|k_0|a\Delta=0.63$ at the disk's inner edge.
}
\end{figure}

\newpage
\begin{figure}
\epsscale{1.0}
\plotone{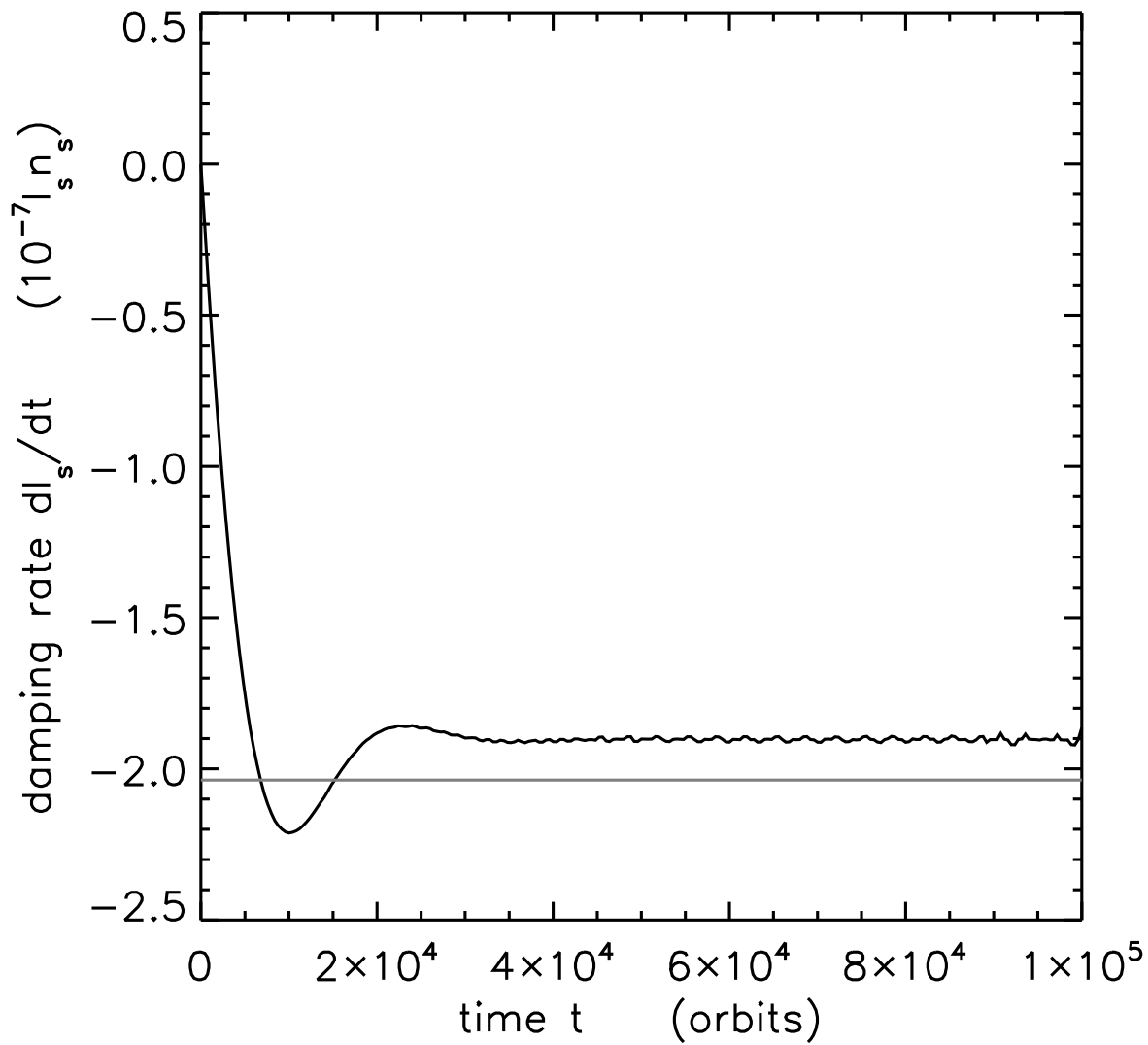}
\figcaption{
  \label{idot}
  The rate at which the satellite launching the wave in
   Fig.\ \ref{Iwave} has its inclination damped,
  $\dot{I}_s$, is plotted versus time t (in units of orbit periods). The solid gray curve
  is the expected rate, from Eqn.\ (\ref{Idamping1}) assuming $|k_0|a\Delta=0.63$
  and $C(|k_0|a\Delta)=0.32$, where $C$ is obtained from Fig.\ \ref{Cfigure}.
}
\end{figure}

\newpage
\begin{figure}
  \epsscale{1.0}
  \plotone{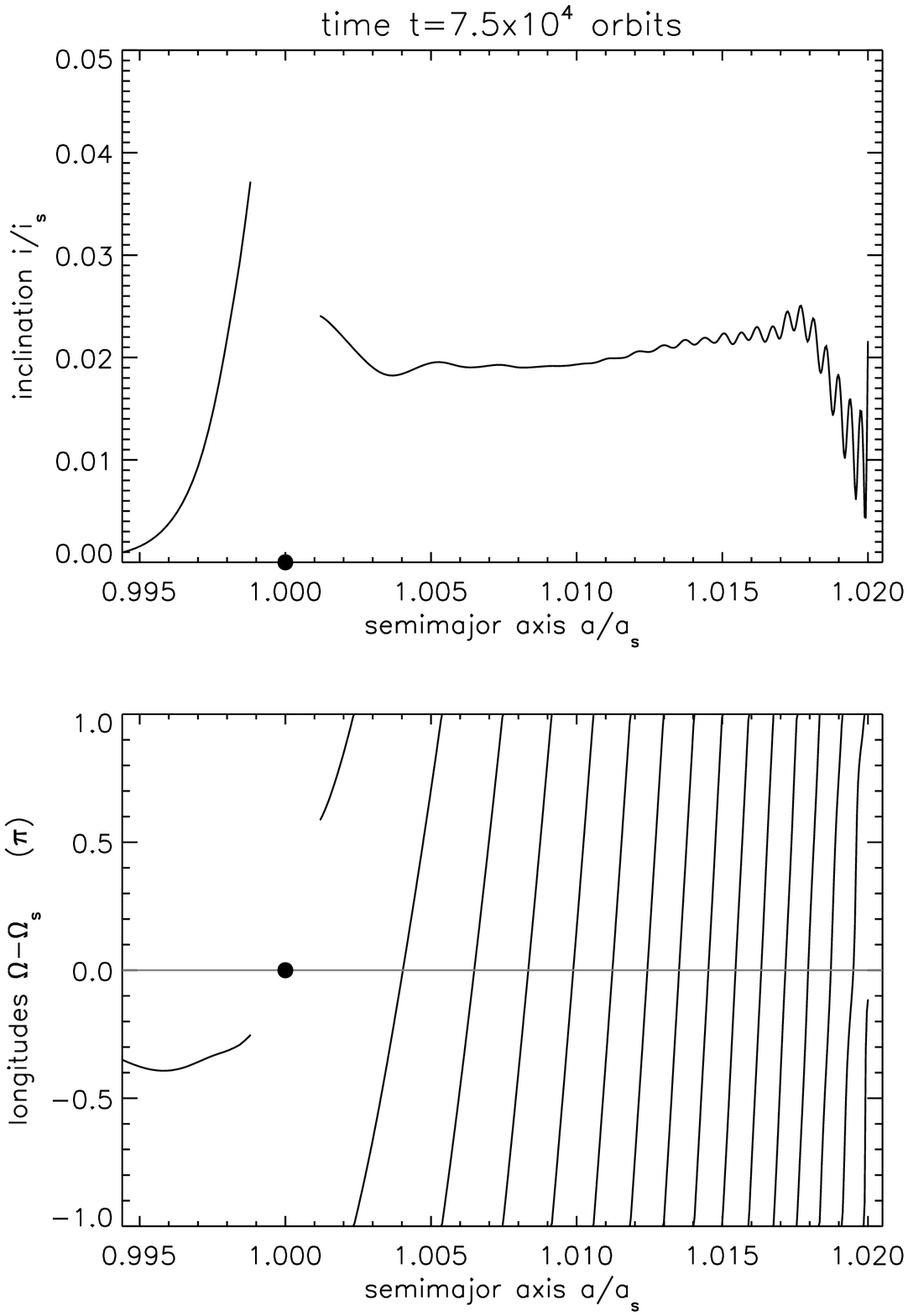}
\end{figure}

\newpage
\begin{figure}
\figcaption{
  \label{pan1}
  A simulation of spiral bending waves launched by an inclined Pan 
  orbiting in the Encke gap in Saturn's A ring.
  Figures show the disk inclinations $I(a)/I_s$ and relative longitudes 
  $\Omega(a)-\Omega_s$ plotted versus semimajor axis $a$ at time 
  $t=7.5\times10^4$ orbits, which is the time it takes the spiral bending wave
  to propagate across the simulated `disk'. Note that
  the simulated disk is actually quite narrow
  since it only extends over $0.99\le a/a_s\le1.02$.
  The system parameters
  are: Pan's mass $\mu_s=8.7\times10^{-12}$ \citep{P05}, semimajor axis
  $a_s=1.34\times10^5$ km, inclination $\sin i_s=10^{-5}$, with an A ring surface density
  $\sigma=50$ gm/cm$^2$ \citep{Retal91} and a normalized disk mass of
  $\mu_d=\pi\sigma a_s^2/M_S=5\times10^{-8}$. The gap half--width
  is $\Delta a_s=160$ km \citep{Betal05}, so its fractional half--width is
  $\Delta=0.0012$. $N=500$ rings are used to
  to simulate the wave in the disk exterior to the satellite, while 50 rings are
  used to resolve the disturbance in the interior disk. The rings' fractional half--widths 
  $\mathfrak{h}$ are set equal to their separations. Saturn's second zonal
  harmonic is $J_2=0.0163$ and the planet's radius is $R_p=0.45a_s$,
  so the system's critical disk mass,
  Eqn.\ (\ref{mu_c}), is $\mu_c=7.8\times10^{-8}$.
}
\end{figure}

\newpage
\begin{figure}
  \epsscale{1.0}
  \vspace*{-5ex}\plotone{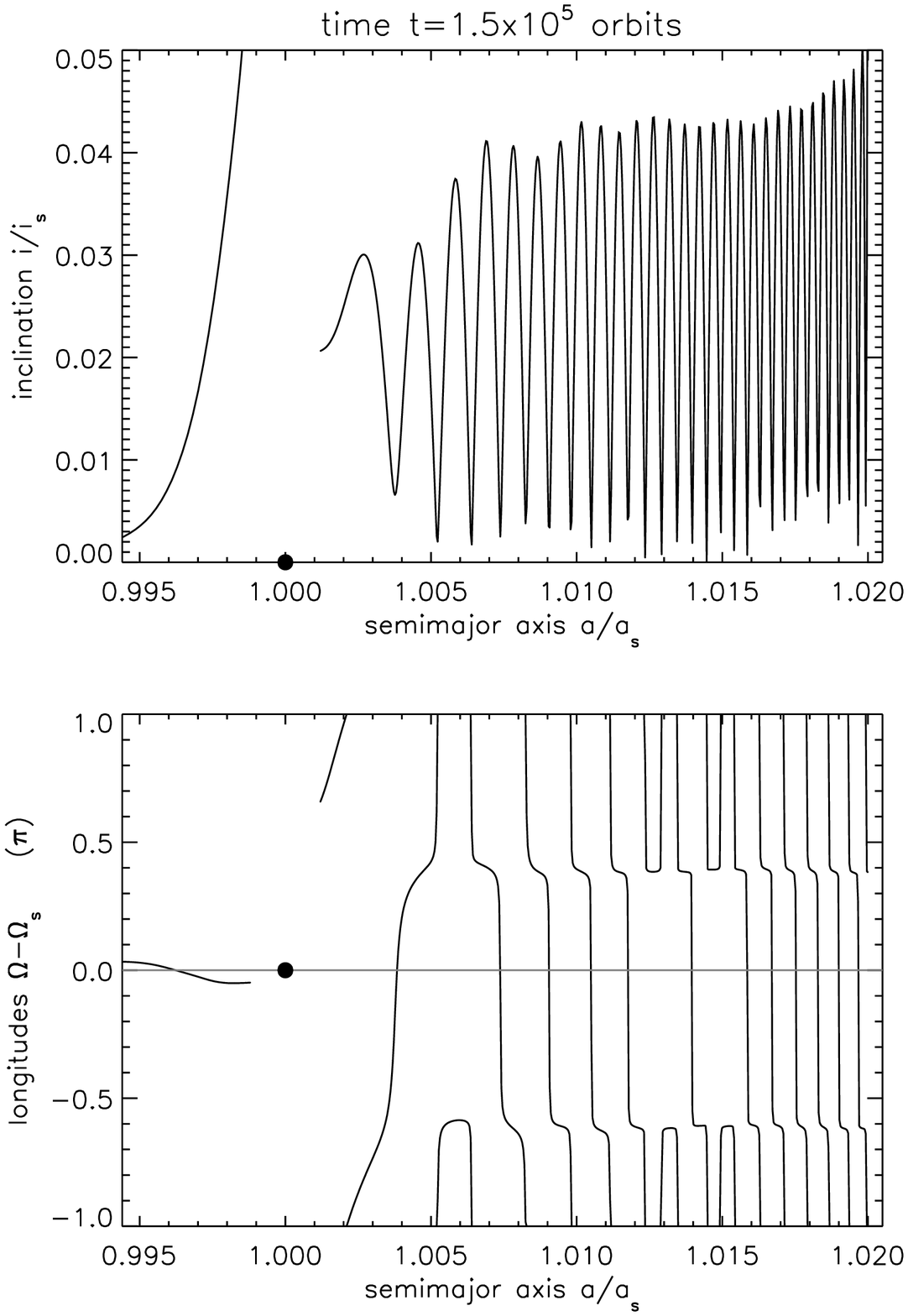}\vspace*{-6ex}
  \figcaption{
    \label{pan2}
    The state of the system described in Fig.\ \ref{pan1} at 
    the later time $t=1.5\times10^5$ orbits,
    which is when the bending wave has reflected at the simulated disk's outer edge
    and returned to the launch site, thereby establishing a standing wave in the disk.
    The fractional error in this system's total angular momentum is
     $|\Delta L_i/L_i|<5\times10^{-5}$.
  }
\end{figure}

\end{document}